\documentclass[aps,prr,amsmath, amssymb, english, ' twocolumn,reprint,floatfix]{revtex4-2}

\usepackage{multirow}
\usepackage{float}
\usepackage[normalem]{ulem}
\usepackage{textgreek}

\usepackage{bbm}
\usepackage{comment}
\usepackage{soul}

\usepackage{tikz}
\usepackage{microtype}
\usepackage{tkz-euclide}
\usepackage{tkz-graph}
\usepackage{epstopdf}
\usepackage{epsfig} 
\usepackage[applemac]{inputenx} 
\usepackage{amsmath}

\usepackage[unicode]{hyperref}
\hypersetup{
 colorlinks,
 citecolor=blue,
 filecolor=blue,
 linkcolor=blue,
 urlcolor=blue
}

\newcommand{\ket}[1]{\left|#1\right>}
\newcommand{\bra}[1]{\left<#1\right|}

\newcommand{\up}{\uparrow}

\graphicspath{{./Images/}}

\date{}
\begin{document}

\title{Dirac points and topological phases in correlated altermagnets}
\author{Lorenzo Del Re} 
\affiliation{Max-Planck-Institute for Solid State Research, 70569 Stuttgart, Germany}
\date{\today} 


\begin{abstract}
We explore a two-dimensional Hubbard model adapted to host altermagnetic states. Utilizing Hartree-Fock (HF) and dynamical mean field theory (DMFT), we uncover that the magnetic solutions of this model feature Dirac points in their spectrum. HF predicts a gap opening at a critical interaction strength, a result corroborated by DMFT calculations at zero temperature. However, at finite temperature and high interaction strengths, Dirac points re-emerge at high energies in the spectral function, even if they are absent in the non-interacting and HF-predicted band structures. Analytical arguments reveal that this phenomenon arises near Mott insulating solutions from the frequency dependence of the self-energy, a behavior not captured by static mean-field theory. 
We identify distinctive dynamical signatures of correlated altermagnetic states in the spin-resolved optical conductivity, notably a double-peak structure likely linked to high-energy Dirac cone-like bands. Moreover, the spin-resolved response exhibits spin-selective activation at different photon energies, pointing to potential applications in spin-dependent optical control.
We also propose perturbations to the model that can induce a topological gap in the spectrum, leading to a transition from topologically trivial to non-trivial states by varying doping, interaction strength, and temperature.
\end{abstract}
\maketitle
\section{Introduction}
Many-electron systems can organise in a multitude of magnetically ordered states by spontaneous symmetry breaking.
Recently a new type of magnetic order i.e. altermagnetism \cite{C5CP07806G,kunes2019,Mazin2021,Mazin2022,Smejkal2022_a,Smejkal2022_b,McClarty2024} has been established as a new category of broken-symmetry phases. Similar to antiferromagnets, altermagnets have a net zero magnetization. However, unlike antiferromagnets, the spin-up and spin-down sub-lattices are not connected by a simple sub-lattice translation. This implies a splitting of spin-up/down bands resulting in spin-resolved anisotropic transport properties with potential application in antiferromagnetic spintronics \cite{Wadley2016,Baltz2018}. 
Even if experiments on real materials have already found good candidates for hosting altermagnetic states \cite{Bai2022,Hariki2024, krempasky2024,zhu2024,jiang2024,regmi2024}, cold atoms trapped in optical lattices represent an alternative route for their realisation. Furthermore, since cold-atomic setups have been employed for the simulation of strongly correlated Mott insulators \cite{jordens2008,esslinger2010,taie2012,tusi2022} and quantum magnetism \cite{hart2015,mazurenko2017,xu2023,shao2024}, they represent the ideal testbed for understanding the interplay between electronic correlations and the emergence of altermangetism in many fermion systems.
\par In this work, we study a two-dimensional version of the Hubbard model suitably modified for hosting altermagnetic states that can be simulated via nowadays available cold-atom technology \cite{Das2024}.
Here, by employing Hartree-Fock (HF) and dynamical mean field theory (DMFT) \cite{georges1996}, we reveal that magnetic solutions of this model host Dirac points in its spectrum and  study their stability as a function of relevant physical parameters such as interaction, doping and temperature.  In particular,  HF predicts a gap opening for a critical value of the interaction strength which is confirmed by our DMFT calculations at zero temperature. At finite temperature and sufficiently strong interaction strength, DMFT predicts the re-emergence of Dirac points at high energy in the spectral function, which are absent in both the non-interacting and HF-predicted band structures.
By means of an approximation valid at strong coupling, we  analytically show that this phenomenon stems from the frequency dependence of the self-energy in the vicinity of Mott insulating solutions and it is not captured by static mean-field theory.
We also investigate the optical conductivity as a probe of the underlying electronic structure in correlated altermagnetic states. Motivated by the presence of high-energy Dirac cone-like features at finite temperature and doping, we analyze the spin-resolved dynamical response to an external electric field. This allows us to identify spectroscopic fingerprints--such as a characteristic double-peak structure in the optical conductivity--that are sensitive to spin polarization and band structure. These features not only provide insights into the many-body physics of the system but also suggest experimentally accessible signatures and possible routes toward spin-selective optical manipulation in strongly correlated materials.
Finally, we propose perturbations to the model that can open a topological gap in the spectrum. Since the topological states \cite{Haldane1988,KaneMele2005,Qi2006,Hasan2010} are inherently connected to the presence of Dirac points in the unperturbed Hamiltonian, the perturbed system undergoes a transition from topologically trivial to non-trivial by varying doping, interaction strength and temperature.


%
\section{The model}
\begin{figure}
    \centering
    \includegraphics[width=0.75\linewidth]{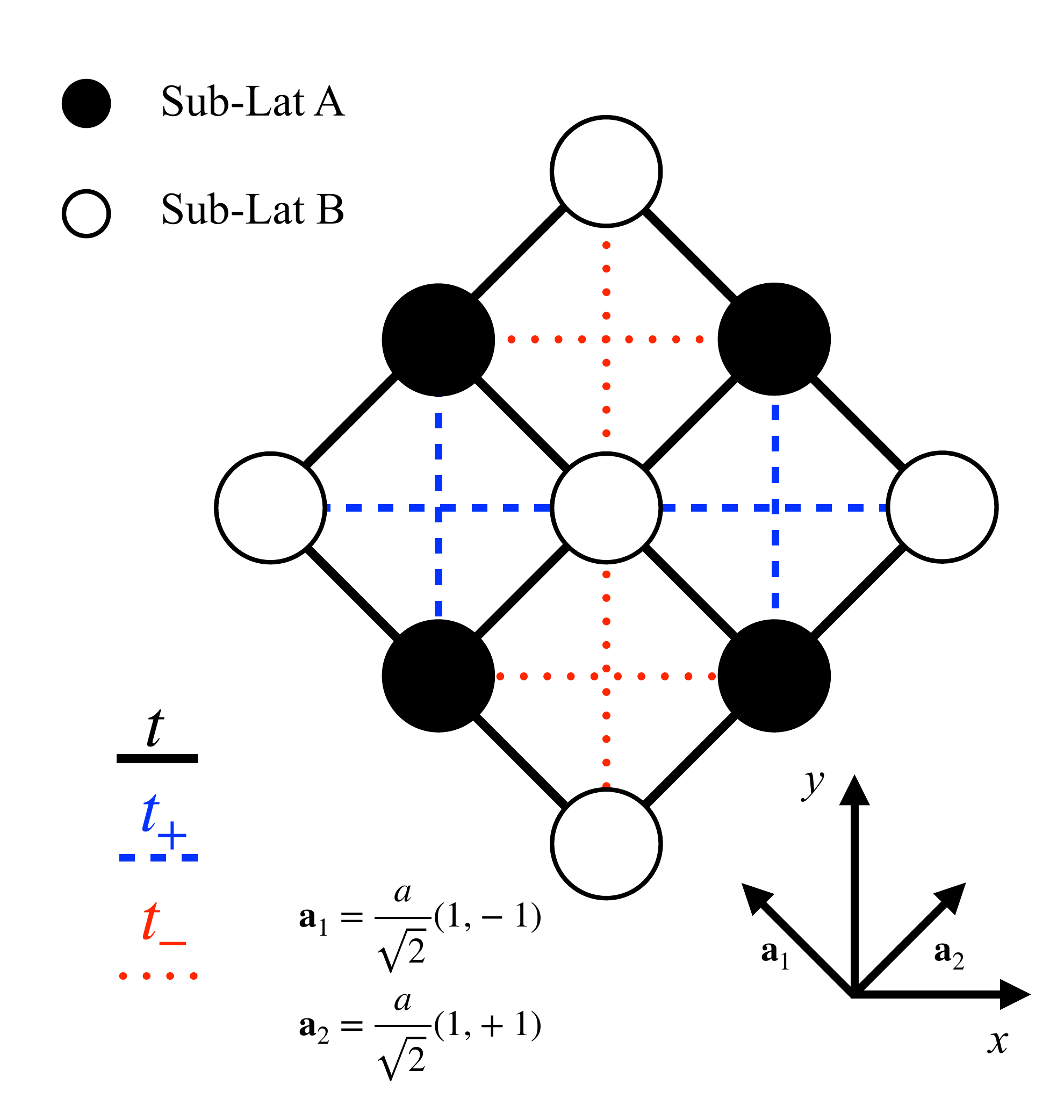}
    \caption{Sketch of the tight-binding model described in Eq.~(\ref{eq:Hubb_model}), where \( t_\pm = t^\prime(1 \pm \delta) \). The reference frame is aligned with the primary ($x$) and secondary ($y$) diagonals. The lattice constant is set to \( a = 1/\sqrt{2} \).
 }
    \label{fig:sketch_tb}
\end{figure}

We consider the Hubbard model in the square lattice with nearest and next-to-the-nearest neighbor hopping (n.n. and n.n.n. respectively):
\begin{align}\label{eq:Hubb_model}
    H &= \sum_{ij\sigma}t_{ij}c^\dag_{i\sigma}c^{\,}_{j\sigma}  + U\sum_{i}\left(n_{i\uparrow}-\frac{1}{2} \right)\left(n_{i\downarrow}-\frac{1}{2}\right),
\end{align}
where the n.n.n. hopping processes depend explicitly on the sub-lattice index and assume different values along the main and secondary diagonals , i.e. $t_{n.n.n} = -t^\prime[1 + (-1)^a\pm \delta]$ for the main (secondary) diagonal where $a = A,B$ is the sub-lattice index \cite{Das2024,PhysRevB.111.104432,he2025altermagnet}, see Figure \ref{fig:sketch_tb}.

Such a hopping process breaks explicitly translational invariance and the system must be addressed by adding sub-lattice indices even in the non-magnetic phase.

\section{Hartree-Fock }
 If we employ a mean-field decoupling of the interactions, the problem is simplified to the following 2$\times$2  Hamiltonian:
\begin{align}\label{eq:eff_MF}
    \mathcal{H}_{\sigma}(k) & = \alpha_k\sigma_0 + \beta_k \,\sigma^{(z)} + \gamma_k \,\sigma^{(x)}, 
\end{align}
where, $\sigma^{(i=x,y,z)}$ are the Pauli matrices, $\sigma_0$ is the 2$\times$2 identity matrix,
$\alpha_k = \frac{f_A(k) + f_B(k)}{2}$, $\beta_k = \frac{f_{A}(k)-f_B(k)-2\sigma\Delta}{2}$, $\gamma_k = \epsilon_k$, $\epsilon_k  = -2t\cos\left(\frac{k_x+k_y}{2}\right) -2t\cos\left(\frac{k_x-k_y}{2}\right)$ being the n.n.  and $f_{a} = -2t^\prime[1 +(-1)^a \delta]\cos({k}_x)-2t^\prime[1 -(-1)^a \delta]\cos({k}_y)$. Here,  $\Delta = U m /2$ is the Hartree term coming from the interaction with $m = n_{A\uparrow} - n_{A\downarrow}$ being the staggered magnetisation.

\subsection{Dirac cones}
The eigenvalues of Eq.(\ref{eq:eff_MF}) read: $\lambda^\pm_{\sigma}(k) = \alpha_k \pm \sqrt{\beta_k^2 + \gamma_k^2}$.
The two bands display a crossing when  $\beta_k$ and $\gamma_k$ simultaneously vanish. 

\begin{figure}
    \begin{center}
    
\includegraphics[width = \columnwidth]{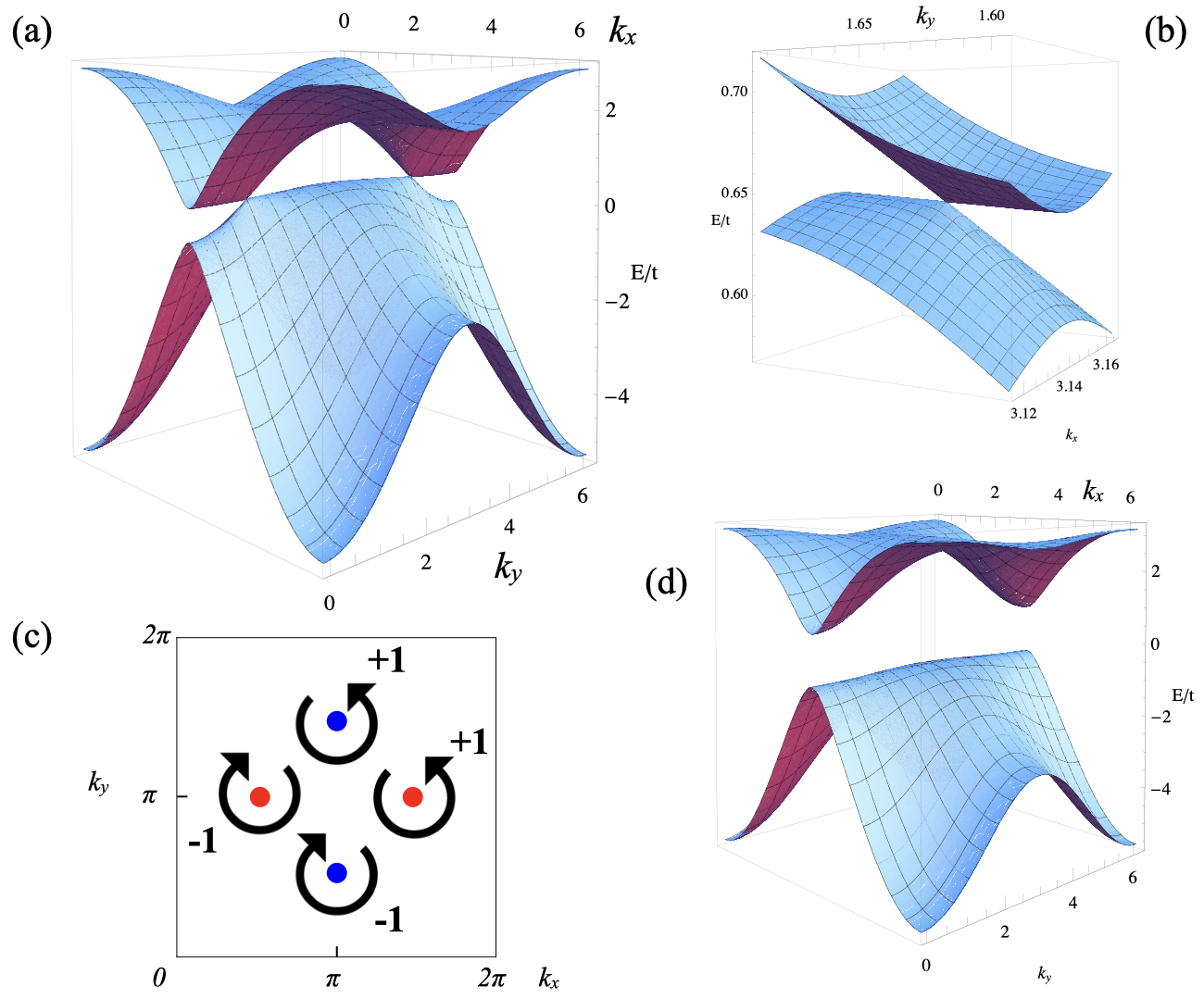}
    \caption{(a) Band structure for $\sigma = \uparrow$ plotted in the Brillouin zone for $t^\prime /t = 0.3$, $\delta = 0.9$ and $\Delta/t = 3/4 < \Delta_c$. (b) Inset of (a) around one of the two crossing points showing the tilted type-I Dirac cone. (c) Positions of the crossing points in the Brillouin zone for $\sigma = \uparrow$ (blue) and $\sigma = \downarrow$ (red) for the same parameter set as in (a). The spin-resolved Dirac points possess opposite winding numbers $W = \pm 1$ (indicated by arrows). (d)  Gapped band structure for $\sigma = \uparrow$ plotted in the Brillouin zone for $t^\prime /t = 0.3$, $\delta = 0.9$ and $\Delta/t = 7/4 > \Delta_c$. }
    \label{fig:bands3D}
    \end{center}
\end{figure}

Assuming $\Delta,t^\prime,\delta > 0$,  the band crossings occur at: 
\begin{align}\label{eq:crossing_points}
\begin{array}{cc}
\displaystyle k_{x} = \pi, k_{y} = \pm\arccos\left(\frac{\Delta}{2t^\prime \delta} -1\right)& \displaystyle  \text{for } \sigma =\, \uparrow \\ \\
\displaystyle  k_{y} = \pi, k_{x} = \pm\arccos\left(\frac{\Delta}{2t^\prime \delta} -1\right)&  \displaystyle \text{for } \sigma =\, \downarrow.
\end{array}
\end{align}
 Let us note that there exists a critical value for the order parameter $\Delta_c = 4t^\prime \delta$ such that when $\Delta < \Delta_c$ the bands display a crossing otherwise they are fully gapped.
In Figure (\ref{fig:bands3D}-a) the band structure  for $t^\prime = 0.3 t$, $\delta = 0.9$ and $\Delta = (3/4) t<\Delta_c$ is shown in the Brillouin zone: there are two crossing points per spin-species where the two bands form tilted Dirac cones \cite{Montambaux2019} (see inset in Figure \ref{fig:bands3D}-b). When $\Delta >\Delta_c$ the two bands are fully gapped as shown in Figure \ref{fig:bands3D}-d.

If the Hamiltonian in Eq.(\ref{eq:eff_MF}) for the $\sigma = \uparrow$ species is expanded around the points defined in Eq.(\ref{eq:crossing_points}), for $\Delta< \Delta_c$, we obtain the following low-energy effective theory:
\begin{align}\label{eq:low-energy}
   \mathcal{H}^\pm_\uparrow&\sim (E_0+\tilde{v}_y q_y )\sigma_0  \pm v_y q_y \sigma^{(z)} + v_x q_x \sigma^{(x)},
\end{align}
where $E_0 = -2t^\prime[\cos(\bar{k}_y)-1]$, $\tilde{v}_y = 2t^\prime\sin(\bar{k}_y)$, $v_x = 2t \cos(\bar{k}_y/2)$, $v_y =  -2t^\prime\delta\sin(\bar{k}_y)$, with $\bar{k}_y = \arccos(\Delta/2t^\prime\delta -1)$, and $\mathbf{q} =(k_x-\pi,k_y\mp\bar{k}_y) $.  

The winding vector of Eq.(\ref{eq:low-energy}) is defined as:

\begin{align}
    \mathbf{W} = \frac{1}{2\pi}\oint_\mathcal{C} \mathbf{n}(\mathbf{q})\times d\mathbf{n}(\mathbf{q}),
\end{align}
where $\mathbf{n}(\mathbf{q}) = \mathbf{h}(\mathbf{q})/|\mathbf{h}(\mathbf{q})|$, with $\mathbf{h}(\mathbf{q}) = (v_x\,q_x,0,\pm v_y\,q_y)$ and $\mathcal{C}$ is a closed curve around the crossing point. The two crossing points possess non-vanishing and opposite winding numbers $\mathbf{W} = (0,\pm 1,0)$ [see Figure \ref{fig:bands3D}]. 

\section{DMFT} In DMFT \cite{georges1996}, the original lattice model is mapped onto an effective Anderson impurity model (AIM) that we solved numerically using exact diagonalisation (ED). As a minimal choice for describing the altermagnetic state, we assume a local self-energy, which is diagonal in sub-lattice and spin indices $\Sigma_{aa^\prime, \sigma \sigma^\prime}(\nu)=\delta_{\sigma\sigma^\prime}\delta_{aa^\prime}\Sigma_{\sigma a}(\nu)$ and has the following property $\Sigma_{\sigma a}(\nu) = \Sigma_{\bar{\sigma}\bar{a}}(\nu)$. Therefore, there are only two independent self-energy components: i.e. $\Sigma_{\up A}(\nu)\equiv \Sigma_\uparrow(\nu)$ and $\Sigma_{\downarrow A}\equiv \Sigma_\downarrow(\nu)$. Given these assumptions the Dyson equation for the $\sigma = \uparrow$ species  reads:
\begin{align}\label{eq:Greens_AB}
    G_\uparrow^{-1}(k,\nu) &=
    \left(
    \begin{array}{cc}
       \zeta_\uparrow(\nu)-f_A(\mathbf{k})  & \epsilon_k \\
        \epsilon_k & \zeta_\downarrow(\nu)-f_B(\mathbf{k})
    \end{array}
    \right),
\end{align}
where $\zeta_\sigma(\nu) = i\nu + \mu - \Sigma_{\sigma}(\nu)$ \cite{georges1996,sangio2006,delre2021_ANA,delre2021AF,reitner2024nonp,delre2025scipost,Giuli2025}.

The self-consistence equation for $\mathcal{G}_0$ is obtained using the inverse Dyson equation, i.e.
   $\mathcal{G}_{0,\uparrow}^{-1}(\nu) = {G}^{-1}_{\text{loc},\uparrow}(\nu) + \Sigma_\uparrow(\nu)$
where ${G}_{\text{loc},\uparrow}(\nu) = \frac{1}{V}\sum_k G_{AA,\uparrow}(k,\nu)$. 
In practice, an initial guess for $\mathcal{G}$ defines the AIM at the first iteration, then the self-energy is extracted using ED, and the Weiss field is updated using the self-consistency condition. The DMFT cycle iterates until convergence of the Weiss field \footnote{See Ref\cite{Capone2007,delre2018,ferraretto2022enhancement} for more technical details on ED-DMFT.}.

\subsection{Paramagnetic solution}
\begin{figure}
    \centering
    \includegraphics[width=0.85\columnwidth]{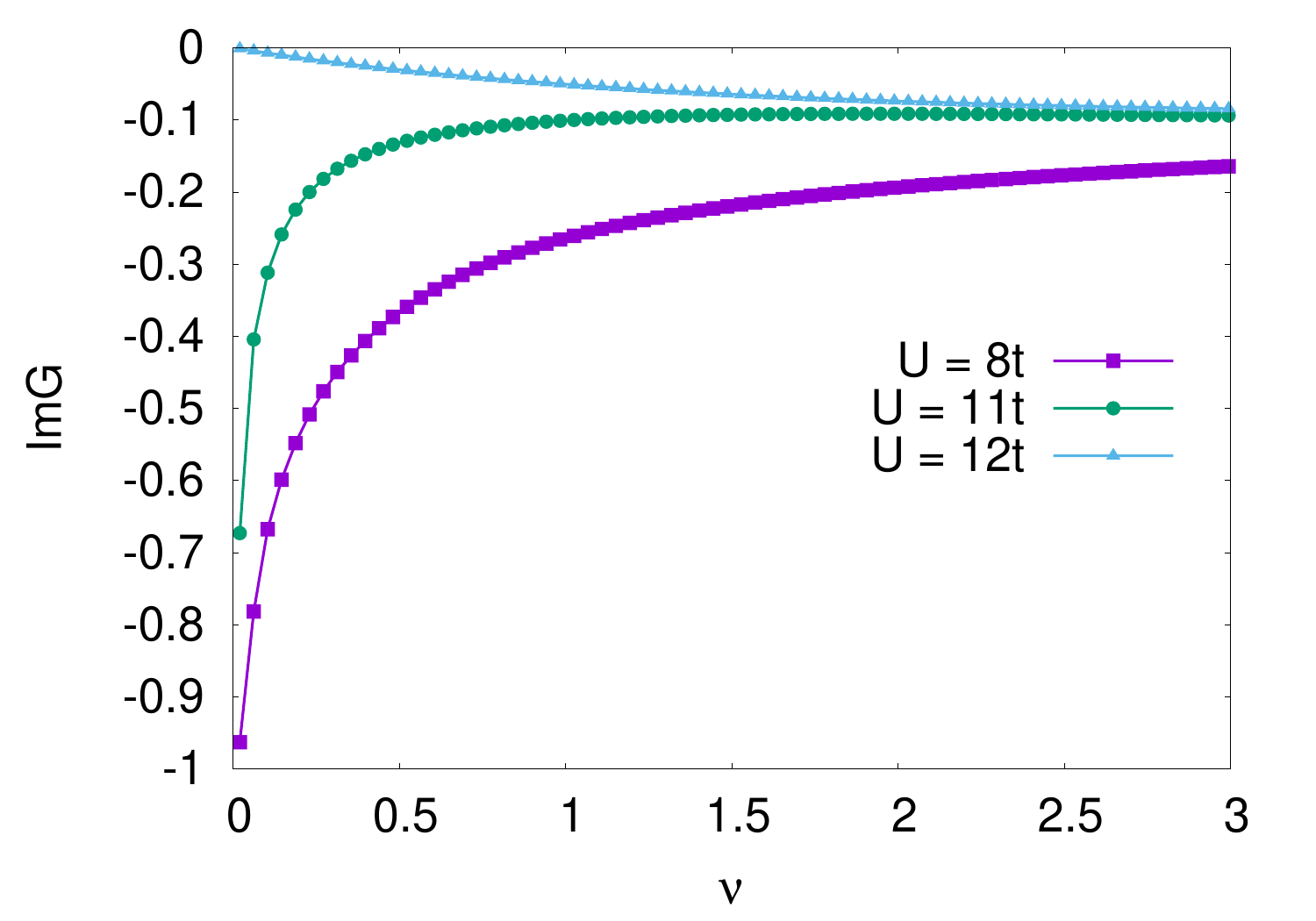}
    \includegraphics[width=0.85\columnwidth]{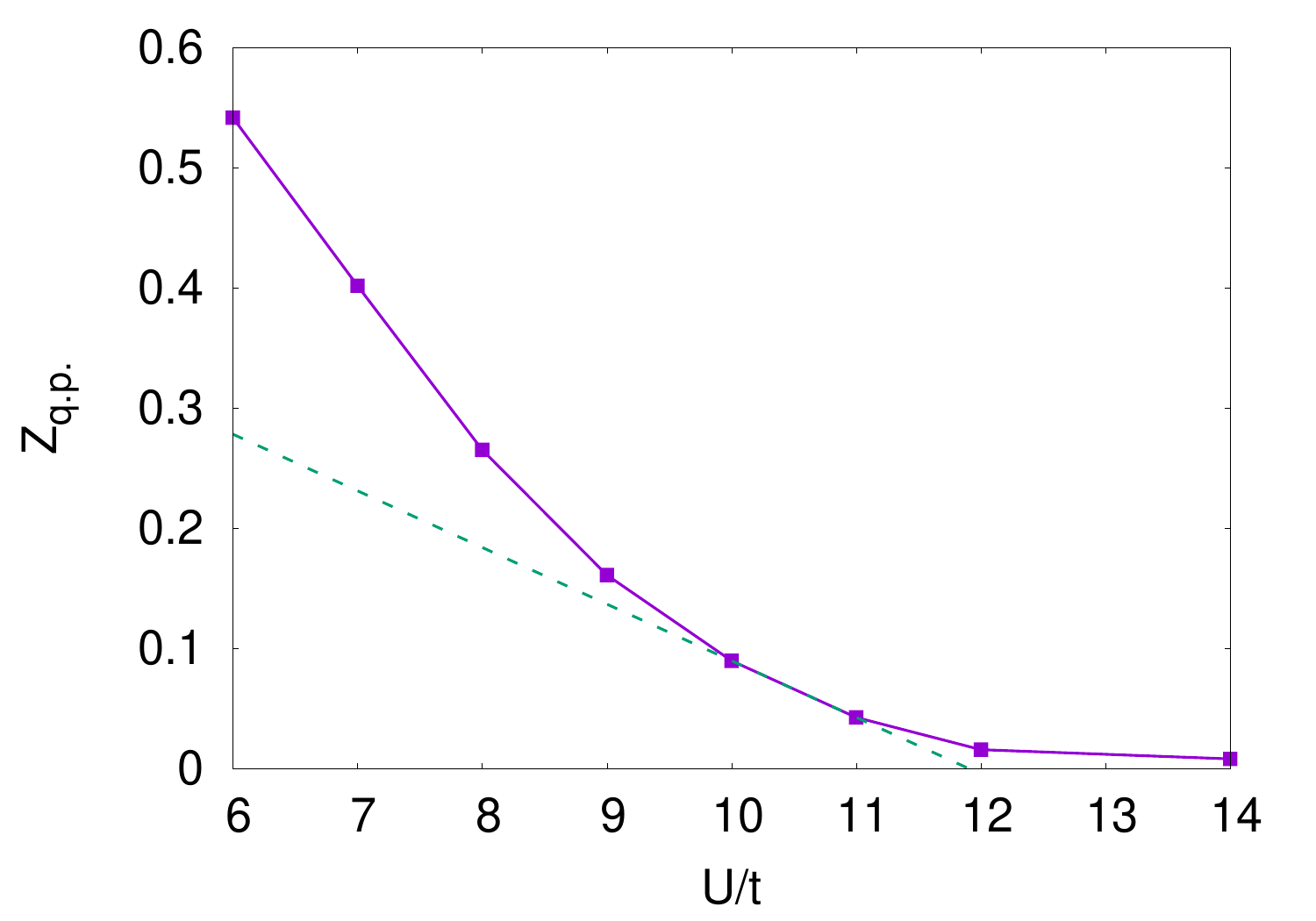}
    \caption{(Top panel) Imaginary part of the electronic Green's function evaluated on the imaginary frequencies $\nu = (2n+1)\pi/\beta$ for three different values of $U/t = 8,\,11,\,12$.
    (Lower panel) Quasi particle weight as a function of on-site interaction at integer filling. The   dashed line has been obtained from a linear extrapolation of the last two metallic solutions at $U/t=10,\,11$. The other parameters for both panels are given by $t^\prime/t = 0.3$, $\delta = 0.9$.}
    \label{fig:PM_data}
\end{figure}

In the symmetric non-magnetic phase $\Sigma_\uparrow = \Sigma_\downarrow$ and the staggered magnetisation vanishes. Within this phase, the system is expected to undergo a metal-to-insulator transition (MIT) upon increasing the onsite interaction at filling $n = 1$.
In the upper panel of Figure (\ref{fig:PM_data}), we show the imaginary part of the Green's function plotted as a function of the Matsubara frequencies $\nu$. For $U < 12 t$, we observe that the imaginary part is finite at the lowest Matsubara frequency, which indicates a metallic behavior. Conversely, for $U = 12\,t$ the imaginary part vanishes linearly in the limit of $\nu \to 0^+$, which signals a vanishing spectral weight at the Fermi level and the opening of a gap.
From the knowledge of the self-energy we can extract the quasi-particle (QP) weight 
$Z_{\text{q.p.}} = \left(1-\partial_\omega \Sigma|_{\omega = 0}\right)^{-1}$, that measures the degree of correlation in the metallic state.  The lower panel of Figure (\ref{fig:PM_data})  shows $Z_{\text{q.p.}}$ as a function of the interaction at fixed density $n = 1$, for $t^\prime = 0.3\, t$  and $\delta = 0.9$.
For this numerical data set and the ones showed in the remainder of this paper  the number of bath sites has been fixed  to $N_b = 5$.
\subsection{Magnetic solution}
\begin{figure}
    \centering
    \includegraphics[width=0.85\columnwidth]{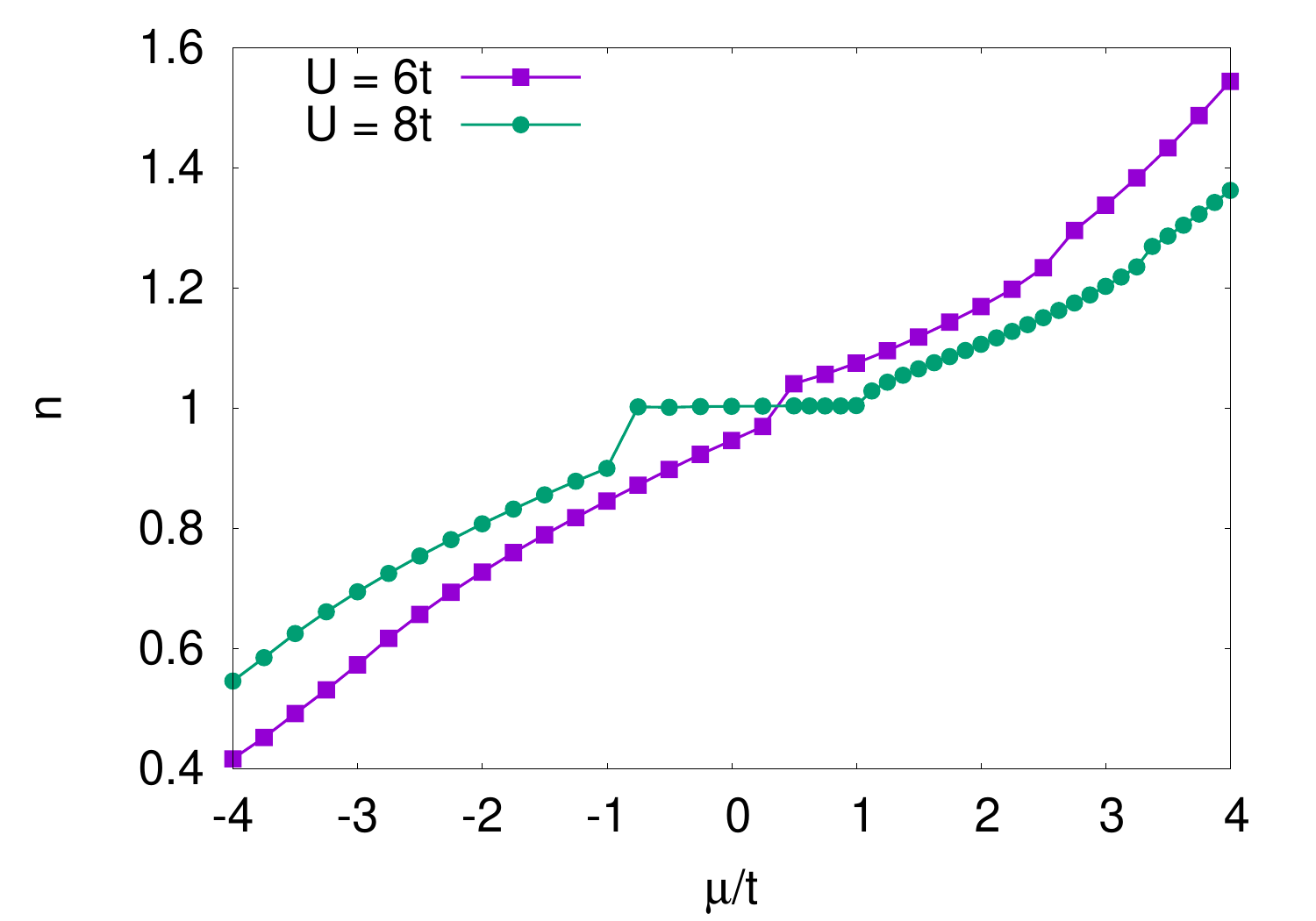}
    \includegraphics[width=0.85\columnwidth]{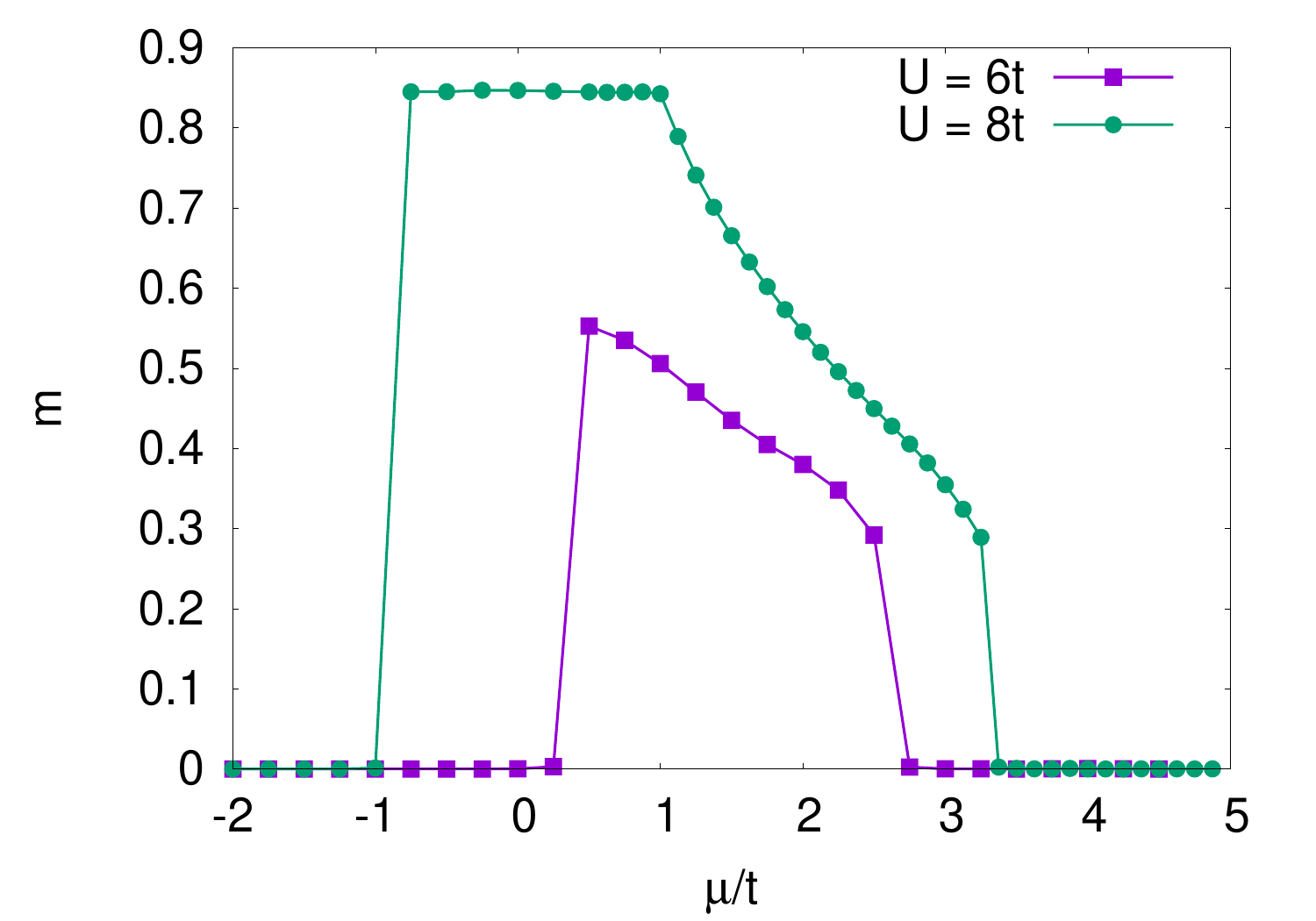}
    \caption{ (Top panel) Density $n = n_{A\uparrow} + n_{A\downarrow}$ as a function of the chemical potential for two values of $U/t = 6,8$, $t^\prime/t =0.3$, $\delta = 0.9$ and zero temperature. (Lower panel) Staggered magnetisation $m = n_{A\uparrow} - n_{A\downarrow}$ as a function of the chemical potential for the same parameters. }
    \label{fig:dens_magn_mu}
\end{figure}
\begin{figure*}
    \centering
    \includegraphics[width=1.8\columnwidth]{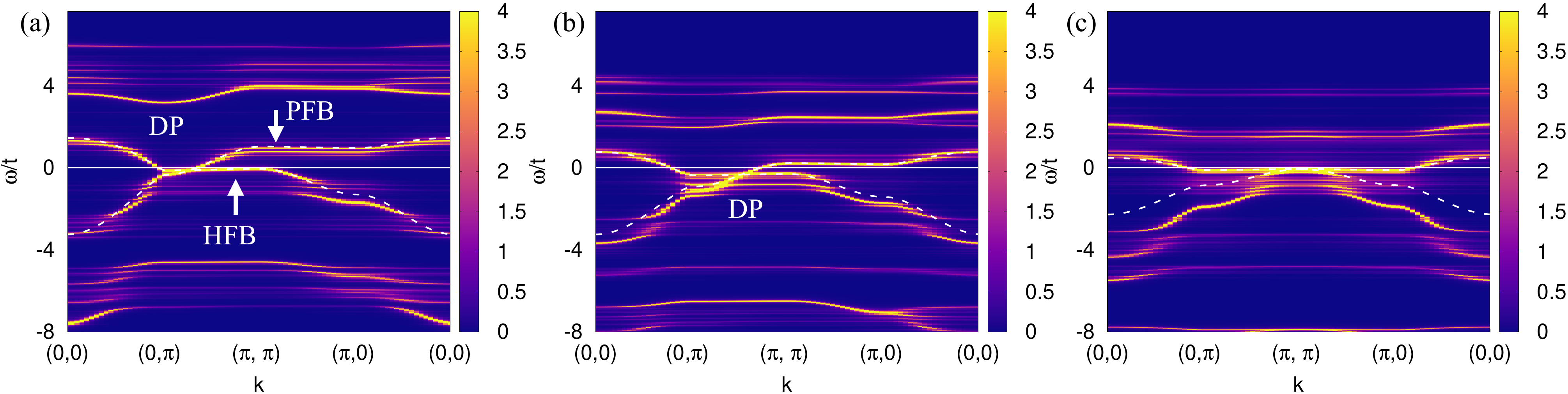}
    \includegraphics[width=1.8\columnwidth]{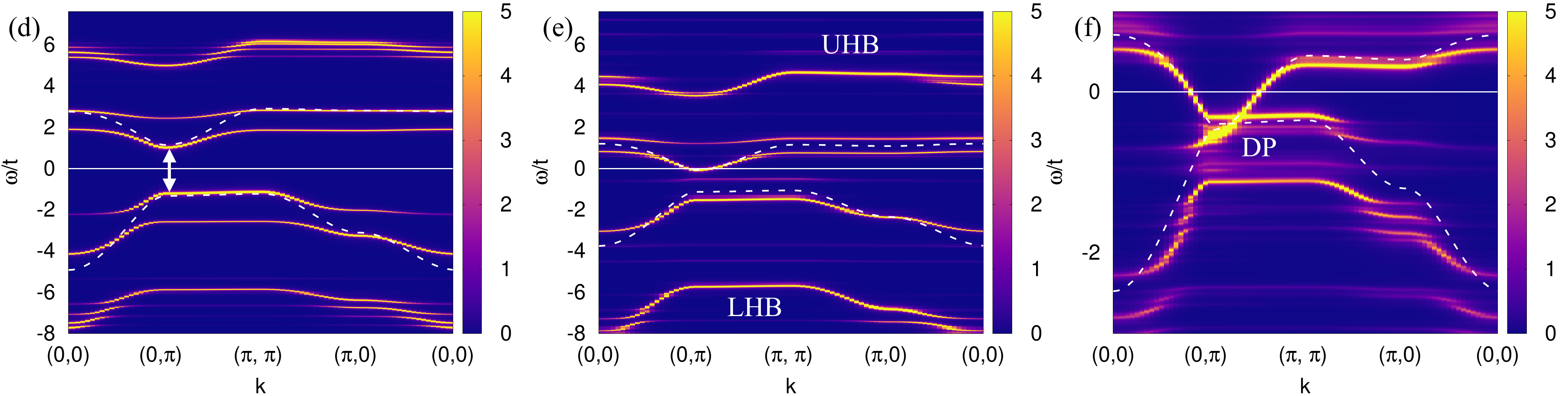}
    \caption{Doping evolution of the spectrum for two different values of the on-site interaction and for $\sigma = \,\uparrow$. (Upper panels) $U/t = 6$, and  $\mu/t = 0.5,\, 2,\, 3$ for (a), (b) and (c) respectively. (Lower panels) $U/t = 8$, and  $\mu/t = 0, \,1.125,\, 2$ for (d), (e) and (f) respectively. Dashed lines are predictions from the QP effective theory defined in Eq.(\ref{eq:eff_hamQP}).}
    \label{fig:spectral_mu}
\end{figure*}
Magnetic solutions are obtained relaxing the constraints imposed by SU(2)-symmetry and letting $\Sigma_\uparrow$, $\Sigma_\downarrow$ assuming independent values.  
We studied the system by varying doping, interaction and temperature and we find three main kind of solutions with broken symmetry: (i) fully gapped insulating AFM without Dirac cones, (ii) a  magnetic metallic phase obtained by doping the fully gapped one without Dirac cones, (iii) a magnetic metallic phase with Dirac cones, (iv) an insulating magnetic phase with high-energy Dirac cones. All these phases display  the particular feature of altermagnets: i.e. a lifted spin degeneracy in the single-particle spectrum.

\subsubsection{Zero Temperature}
In Figure(\ref{fig:dens_magn_mu}), we show the density $n = n_{A\uparrow} + n_{A_\downarrow}$ and staggered-magnetization $m = n_{A\uparrow} - n_{A_\downarrow}$ curves as a function of the chemical potential for two different values of $U/t = 6,8$ at zero temperature and for $t^\prime/t = 0.3$ and $\delta =0.9$. For $U = 6 t$, the density is a continuously increasing function of the chemical potential and the system is always found in a metallic state for this value of the interaction. The staggered-magnetisation is peaked close to integer filling $n = 1$ and it decreases continuously upon particle doping, whereas it vanishes abruptly for hole doping.

At $U = 8 t$, the density curve displays a plateau for $n = 1$ which signals the onset of an insulating phase at integer filling. Also in this case, the staggered-magnetisation jumps to zero in a discontinuous fashion upon hole doping. For moderate particle doping the system becomes a magnetic metal and the order parameter decreases to intermediate values before vanishing. 


Further insights about the nature of these metallic and insulating states can be grasped by inspection of the single-particle spectral function $A_\sigma(k,\omega) = -\text{Im}\text{Tr}\, G_\sigma(k,\omega + i0^+)$, which is obtained from the analytic continuation of Eq.(\ref{eq:Greens_AB}), that in ED can be obtained analitically via Lehmann representation.

The top panels of Figure \ref{fig:spectral_mu} show the evolution of the single-particle spectrum as a function of doping for $U/t = 6$ and $\sigma = \uparrow$ plotted as an intensity map in the plane $\omega$-vs-$k$.  From left-to-right, the chemical potential is set to $\mu/t= 0.5,2,3$.
At $\mu =0.5$, that corresponds to $n\sim 1$, the order parameter assumes its maximal value shown in Figure \ref{fig:dens_magn_mu} (blue curve). We observe that the spectral function displays a linear crossing in the direction $(\pi,\pi)-(\pi,0)$, where, according to Eq.(\ref{eq:crossing_points}), we expect to find a Dirac cone. Furthermore, the spectrum is characterised by the presence of two quasi-flat bands one at higher energy, which we refer to as particle-flat-band (PFB), and the other at lower energy, that we named hole-flat-band (HFB). 
At $\mu/t = 2$, the chemical potential (displayed as a solid white horizontal line) lies below the PFB  and the Dirac point is still visible.
By further increasing the chemical potential, the altermagnetic state becomes unstable and the Dirac point disappears, as shown in Figure (\ref{fig:spectral_mu} c).
We observe that at $\mu/t =0.5$ the chemical potential lies slightly above the HFB and by decreasing doping the order parameter vanishes, as shown in Figure (\ref{fig:dens_magn_mu}).

The lower panels of Figure \ref{fig:spectral_mu} display  the spectral function at $U/t = 8$ for different values of the Fermi level. We note that, at $\mu = 0$, which corresponds to integer filling (see Figure \ref{fig:dens_magn_mu}), the spectrum is gapped, with a  gap of $\tilde{\Delta} \sim 2t$ that coincides with the plateau width of the density curve in Figure \ref{fig:dens_magn_mu}. We note that the spectral weight redistributes to higher positive and negative energies forming the Hubbard bands. At $\mu/t = 1.125$ [Figure (\ref{fig:spectral_mu} e)], the Fermi level is high enough to populate the first low-energy band and the system becomes metallic. However, differently from the AF-metallic phase at $U/t =6$ the spectrum does not feature any Dirac points and it similar to the one of a doped Mott-insulator. Upon further increasing doping, at $\mu/t = 2$
Dirac cones arise in the spectrum, as shown in Figure (\ref{fig:spectral_mu} f). For greater values of $\mu/t$ where the symmetric phase is restored, the Dirac points vanish again (not shown).

\begin{figure*}[htbp]
    \centering
    \includegraphics[width=1.8\columnwidth]{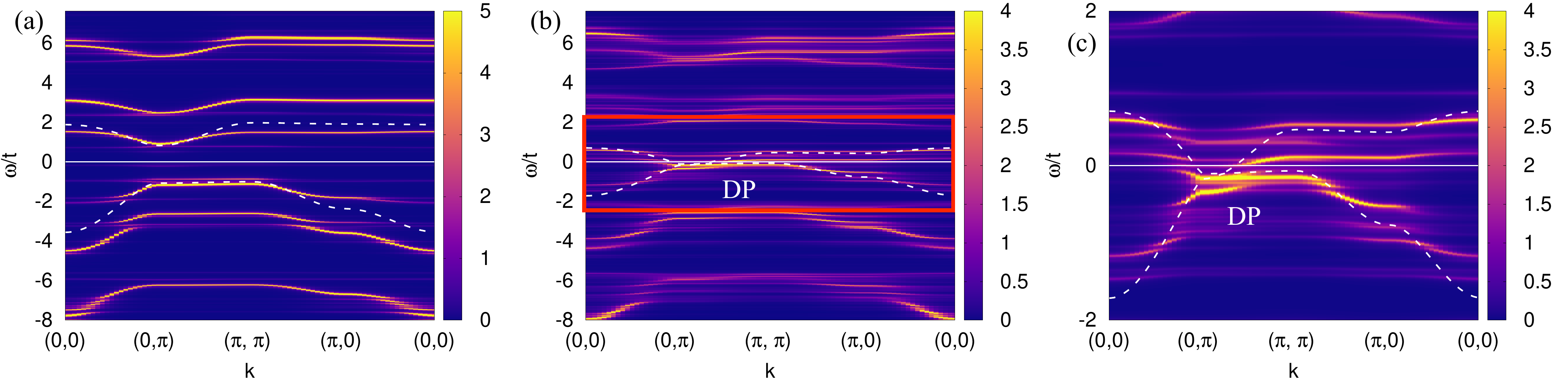}
    \includegraphics[width=1.8\columnwidth]{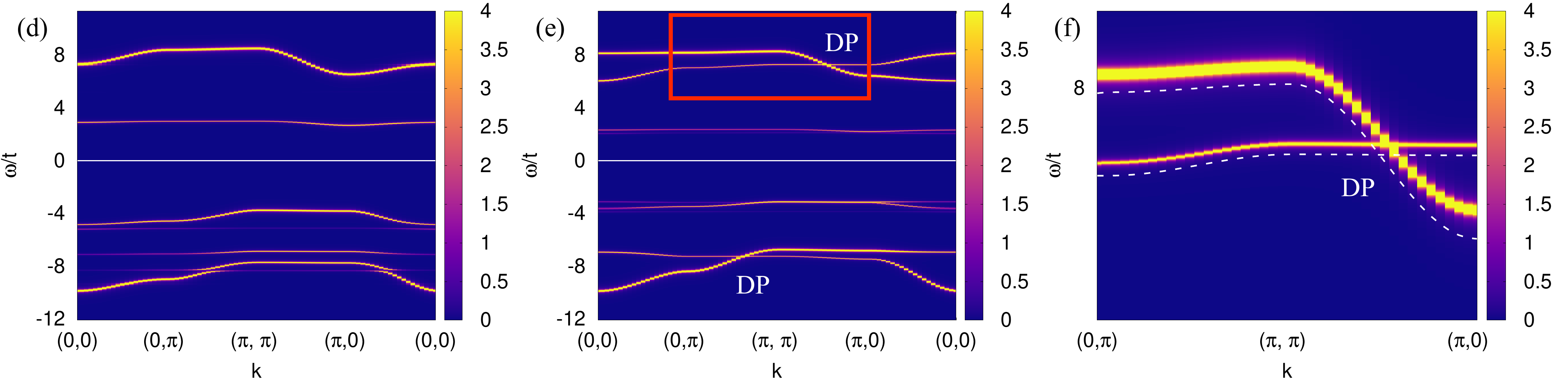}
    \caption{\small Temperature evolution of the spectrum for two different values of the interaction and for $\sigma = \,\uparrow$. (Upper panels) $U/t = 8$ and $\beta t = 5, \,3.425 $ for (a) and (b) respectively. (c) is an inset of (b). The dashed white lines are the results of the QP effective theory defined in Eq.(\ref{eq:eff_hamQP}). (Lower panels) $U/t = 14$ and $\beta t = 25,\, 9$ for (d) and (e) respectively. (f) is an inset of (e). The dashed white lines are the results of the strong coupling expansion based on the approximation presented in Eq.(\ref{eq:AL}) (see main text). }
    \label{fig:spectr_FT}
\end{figure*}

In order to rationalise the DMFT numerical data it is useful to derive an effective low-energy theory. For this purpose we consider a linear expansion in frequency of the self-energy, i.e. $\Sigma_{\sigma}(\nu) \sim \Sigma_{0\sigma} + \alpha_{\sigma} i\nu $, where $\Sigma_{0\sigma} = \text{Re}\Sigma_{\sigma}(\nu \to 0^+)$ and $\alpha_\sigma = \partial_\nu \text{Im}\Sigma_{\sigma}(\nu)|_{\nu \to 0^+}$. Once this simplified expression of $\Sigma$ is substituted into Eq.(\ref{eq:Greens_AB}), we obtain the following approximation for the Green's function: ${G}^{-1}_\sigma(k,\nu) \sim Z^{-1/2} \cdot \left(i\nu - \tilde{\mathcal{H}}_{k\sigma}\right)\cdot Z^{-1/2}$, where
\begin{align}\label{eq:eff_hamQP}
\tilde{\mathcal{H}}_{k\sigma} &= Z^{1/2}\cdot(\mathcal{H}_{0\sigma}(k) + \Sigma_0) \cdot Z^{1/2}, 
\end{align}
is the effective QP Hamiltonian, ``$\cdot$" represents matrix multiplication, $\Sigma_0 = \text{diag}(\Sigma_{0\uparrow},\Sigma_{0\downarrow})$ represents the Hartree term, 
$Z = \text{diag}(Z_\uparrow,Z_\downarrow)$ with $Z_{\sigma} = (1-\alpha_\sigma)^{-1}$ being the spin/sub-lattice resolved QP renormalisation factor.
If we neglect the spin dependence of the QP weight, the effective Hamiltonian in Eq.(\ref{eq:eff_hamQP}) coincides with the one defined in Eq.(\ref{eq:eff_MF}) times a pre-factor and where $2\Delta = {\Sigma_{0\uparrow} - \Sigma_{0\downarrow}}$. The bands obtained from diagonlising the 2$\times$2 matrix in Eq.(\ref{eq:eff_hamQP}) are shown as white dashed lines in Figure \ref{fig:spectral_mu}  and they well capture the salient features of the spectral function at low energy. At zero temperature,  the Hartree term $2\Delta = \Sigma_\uparrow-\Sigma_\downarrow$ decreases by increasing (decreasing) doping and a Dirac cone at low-energy is formed when it becomes lower than the critical value $\Delta_c =4 t^\prime \delta$, as we checked  numerically. 

\subsubsection{The role of temperature}
 Since temperature lowers the order parameter, one might expect that it could favor the appearance of Dirac cones in the energy spectrum, similarly to what we have observed at zero temperature as a function of doping.

The upper panels of Figure \ref{fig:spectr_FT} show the temperature evolution of the spectrum for $U/t = 8$. Indeed, we observe that, since the Hartree term decreases with temperature,  when $\Delta < \Delta_c$ a Dirac cone appears close to the Fermi level and its behavior is well described by the low-energy theory. 

However, for higher values of the interaction strength, where a paramagnetic Mott insulating solution already exists,  the  temperature evolution of the spectrum  changes qualitatively even in the broken symmetry phase. This is shown in the lower panels of Figure \ref{fig:spectr_FT} for $U/t = 14$: at low temperature  Figure (\ref{fig:spectr_FT} d) the spectrum is fully gapped and no Dirac cones appear. 
However, for intermediate values of temperature  Figure (\ref{fig:spectr_FT} e), we observe that even if the system is still gapped, two Dirac points emerge at high negative and positive energies. 
These features are completely missed by the effective QP-theory and they never appear in mean-field theory. 

An alternative approximation to the QP effective theory, which fails  at  intermediate temperature where dynamical correlations are not negligible anymore,  is obtained by evaluating the self-energy in the atomic limit of Eq.(\ref{eq:Hubb_model}) in the presence of a finite  magnetisation, that reads:
\begin{align}\label{eq:AL}
    \Sigma^{\text{a.l.}}_\sigma(z) &= -\frac{U}{2}\,\frac{\sigma \, m \, z - \frac{U}{2}  }{z - \sigma \, m\,\frac{U}{2}},
\end{align}
where $z$ represents the frequency in the complex plane. 
If we substitute Eq.(\ref{eq:AL}) into Eq.(\ref{eq:Greens_AB}) we obtain an approximated version of the Green's function that we expect to be valid for large enough interaction strengths. The Green's function poles, which determine the structure of the single-particle spectrum, are given by the zeros of the determinant of Eq.(\ref{eq:Greens_AB}), that by fixing $k_x = \pi$ or $k_y = \pi$ simplifies  as following:
\begin{align}\label{eq:det}
    \left(z - f_A(k) - \Sigma^{\text{a.l.}}_\uparrow(z)\right)\left(z - f_B(k) - \Sigma^{\text{a.l.}}_\downarrow(z)\right) = 0\,.
\end{align}
Within this approximation, at $T = 0$, the system is fully polarised and $m = 1$. We observe that, when this limit is considered the self-energy becomes a real constant $\Sigma^{\text{a.l.}}_{\sigma}(z, T = 0) = -\sigma \frac{U}{2}$ and the system is weakly correlated. In this case, Eq.(\ref{eq:det}) admits two solutions $z_{\uparrow(\downarrow)} = \mp \frac{U}{2} + f_{A(B)}(k)$: the spectrum is fully gapped and the two bands never cross.

However, at finite temperature $m<1$, the self-energy starts to acquire a non-trivial frequency dependence and Eq.(\ref{eq:AL}) admits four solutions that, when $U \gg f_{A(B)}$, read: $z^\pm_{\uparrow} = \frac{1 \mp m}{2} f_{A}  \pm\frac{U}{2}$ and  $z^\pm_{\downarrow} =\frac{1 \pm m}{2} f_{B}  \pm\frac{U}{2} $. Crossing for the lower (upper) bands occur at $k_{y (x)} = \pi$ and $k_{x(y)} = \pm\arccos\left(\frac{m-\delta}{m+\delta}\right)$.
Hence, for large values of $U$ there are two pairs of Dirac points per spin located at different energies. 
The visibility of the crossing is related to the intensity of the new solutions brought by the frequency dependence of the self-energy for $m<1$, which for the lower (upper) branches  is calculated as the residue of the Green's function at $\overline{z} = z^{+(-)}_{\uparrow (\downarrow)}$, that amounts approximately to Res$\left.G_{AA,\uparrow}\right|_{z = \overline{z} } \sim \frac{1-m}{2}$.  Therefore, at strong coupling, we expect to observe the Hubbard bands crossing more clearly at intermediate temperature/magnetisation values, which is confirmed by our numerical calculations.

\section{Optical conductivity}
Here, we analyze the optical conductivity of the system across different parameter regimes in the limit where the interaction strength is much larger than the hopping amplitude. We consider the system's response to an electric field, introduced as the time derivative of a homogeneous, time-dependent vector potential \( \mathbf{A}(t) \). This vector potential enters the model via the Peierls substitution, implemented by modifying the hopping term as \( t_{ij} \to t_{ij} e^{i(\mathbf{R}_i - \mathbf{R}_j) \cdot \mathbf{A}(t)} \). Using the Kubo formalism, we derive the following expression for the spin-resolved optical conductivity:
\begin{align}\label{eq:conduct}
    \text{Re}\,\sigma^{\mu\nu}_\sigma(\omega + i\eta) = \frac{1}{V}\sum_\mathbf{k}\int_{-\infty}^{+\infty} d\omega^\prime\Lambda_{k\sigma}^{\mu\nu}(\omega,\omega^\prime)  F(\omega,\omega^\prime) ,
\end{align}
where 
$\Lambda_{k\sigma}^{\mu\nu}(\omega,\omega^\prime) = \text{Tr} \,\left[V_k^\mu \cdot \mathcal{A}^\sigma_{k}(\omega^\prime)\cdot V_k^\nu \cdot \mathcal{A}^\sigma_{k}(\omega^\prime + \omega)\right] $, with $\mathcal{A}^\sigma_{k}(\omega) = -\text{Im}G_\sigma(k,\omega + i\eta)$ is the 2$\times$2 spectral function matrix, $V_k^\mu = \nabla \mathcal{H}_0(\mathbf{k})\cdot \hat{e}_\mu$ is the velocity matrix along the $\mu$ direction and $F(\omega,\omega^\prime) = \frac{f(\omega^\prime)-f(\omega +\omega^\prime)}{\omega}$, with $f$ being the Fermi-Dirac distribution function. 
The charge current induced by an electric field is proportional to the sum over spin indices of the quantity in Eq.~(\ref{eq:conduct}), whereas the spin current is proportional to their difference. Due to symmetry considerations, the off-diagonal components of the conductivity tensor (\( \mu \ne \nu \)) vanish. However, the diagonal components remain spin-dependent, with \( \sigma^{\mu\mu}_\sigma \ne \sigma^{\mu\mu}_{\overline{\sigma}} \), and satisfy the relation \( \sigma^{xx}_\sigma = \sigma^{yy}_{\overline{\sigma}} \).

\begin{figure}
    \centering
    \includegraphics[width=0.8\linewidth]{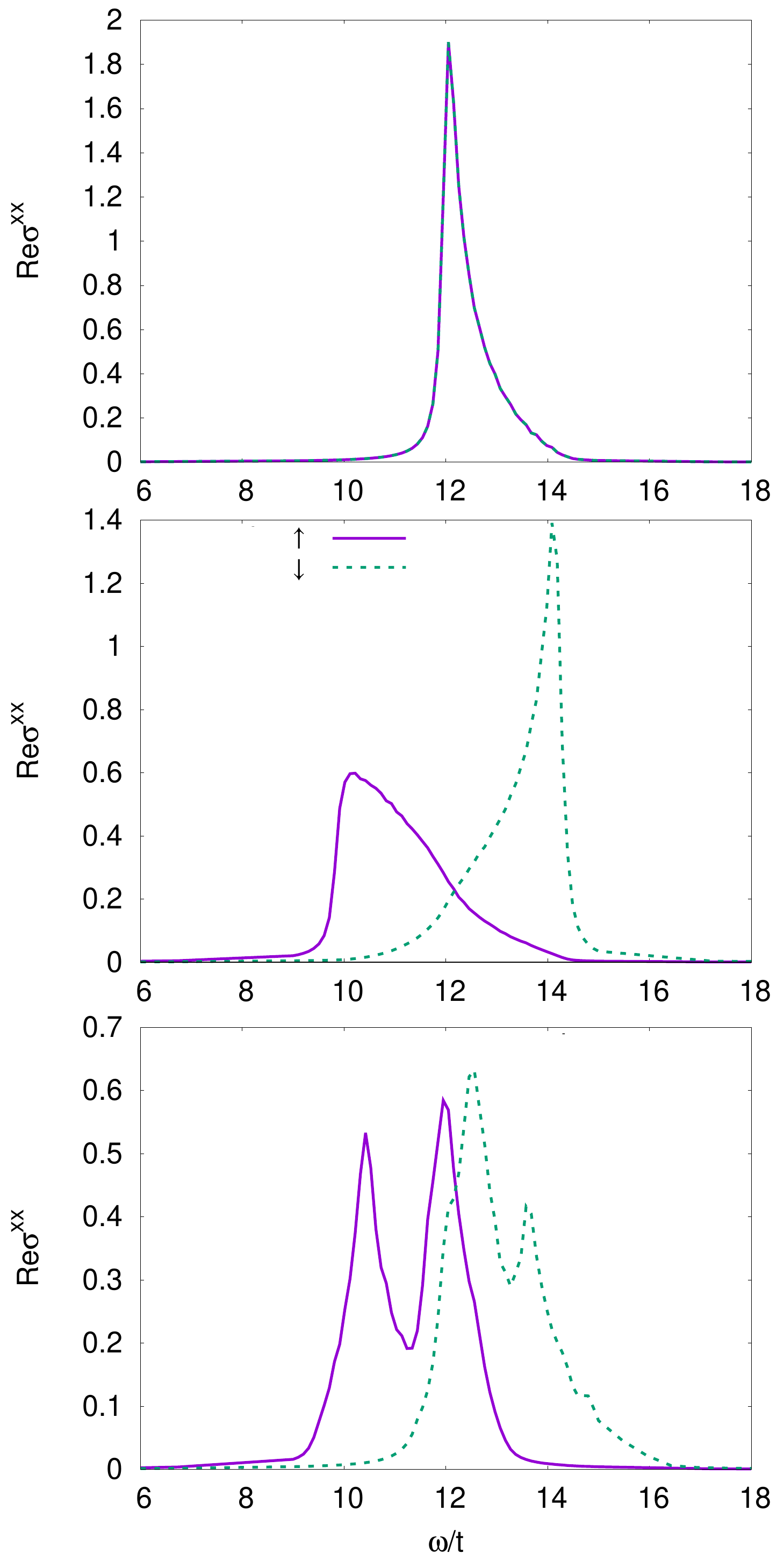}
    \caption{Spin-up (down) optical conductivity is shown as a thick blue (red) line at $U/t = 12$ and $t^\prime/t = 0.3$ for three different cases.
Top panel: $m = 1$, $\delta = 0$ -- fully polarized standard antiferromagnet.
Center panel: $m = 1$, $\delta = 0.9$ -- altermagnetic ground state at $T = 0$.
Lower panel: $m = 0.5$, $\delta = 0.9$ -- correlated altermagnet at finite temperature.}
    \label{fig:opt_cond}
\end{figure}

The static limit of Eq.~(\ref{eq:conduct}) has already been studied in the metallic phase in Ref.~\cite{Das2024}. In this section, we focus on the dynamical structure of the optical conductivity in the insulating phase. To this end, we evaluate Eq.~(\ref{eq:conduct}) numerically using the self-energy expression given in Eq.~(\ref{eq:AL}), which is appropriate in the strong-coupling regime. Specifically, we set \( U/t = 12 \), \( t'/t = 0.3 \), and \( \beta t = 10 \). For the numerical evaluation of Eq.~(\ref{eq:conduct}), we used an \( 80 \times 80 \) \(\mathbf{k}\)-point grid and fixed \( \eta = 0.05 \). We analyze three distinct physical scenarios:

The first case (upper panel of Fig.~\ref{fig:opt_cond}) corresponds to the response to an electric field in a standard antiferromagnetic ground state, obtained by setting $\delta = 0$ and $m = 1$. In this case, the optical conductivity spectrum is peaked around $\omega = U$, as expected for a Mott insulating phase. 
We also note that the spin-up and spin-down contributions are identical due to Kramers degeneracy.

The second case (central panel of Fig.~\ref{fig:opt_cond}) corresponds to a fully polarized altermagnetic phase, obtained by setting $\delta = 0.9$ and $m = 1$. In this regime, the spin-resolved optical conductivity displays markedly different behavior for the two spin species. Each spectrum is characterized by a single-peaked function, with the maximum located below the interaction energy $U$ for spin-up and above $U$ for spin-down.

The third case (lower panel of Fig.~\ref{fig:opt_cond}) corresponds to a correlated altermagnetic state with intermediate polarization, obtained by setting $m = 0.5$ and $\delta = 0.9$. As discussed in the previous section, the spectral function in this phase features two additional bands crossing the original ones present at full polarization, forming two high-energy Dirac cones located at distinct energies.
These more intricate spectral features are reflected in the spin-resolved optical conductivity, which exhibits a characteristic double-peak structure with a pronounced minimum in between.

Let us finally note that our calculations are valid within the DMFT approximation, which assumes a local self-energy and a local vertex function. In the presence of a momentum-dependent vertex function, one would need to include vertex corrections in the expression for the conductivity, which would likely lead to quantitative modifications of the results reported here \cite{PhysRevB.94.085110}. This would considerably complicate the calculation and is beyond the scope of the present work; we leave this for future study.
\section{Topological phase transition}
 Adding an appropriate perturbation to the free Hamiltonian it is possible to obtain bands carrying non-trivial topology \cite{parshukov2024,menghan2025dirac,PhysRevLett.134.096703}. Here we shall consider two possibilities: (i) a spin-independent and (ii) spin-dependent complex hopping. In case (i) $\delta \mathcal{H}_\sigma = \sigma^{(y)} g_k$, where $g_k = -2t_c\cos(\frac{k_x -k_y}{2})$. Hence, the Hamiltonian defined in Eq.(\ref{eq:eff_MF}) transforms into $\mathcal{H}_\sigma +\delta \mathcal{H}$ and it has two gapped bands with Chern numbers $\mathcal{C}_{\sigma}^\pm = \pm\text{sign}(m\,v_x\, v_y)$, where the massive term reads: $m = -2t_c\sin(\overline{k}_y/2)$ [see Appendix \ref{sec:appendix}].

 In case (ii) we set $\delta \mathcal{H}_\sigma = \sigma^{(y)}  \sigma g_k$ and since the hopping sign  is spin-dependent the spin-resolved Chern number is  now given by $\mathcal{C}_{\sigma}^\pm = \pm\sigma\text{sign}(m\,v_x\, v_y)$.

\begin{figure}
    \centering
    \includegraphics[width=0.85\columnwidth]{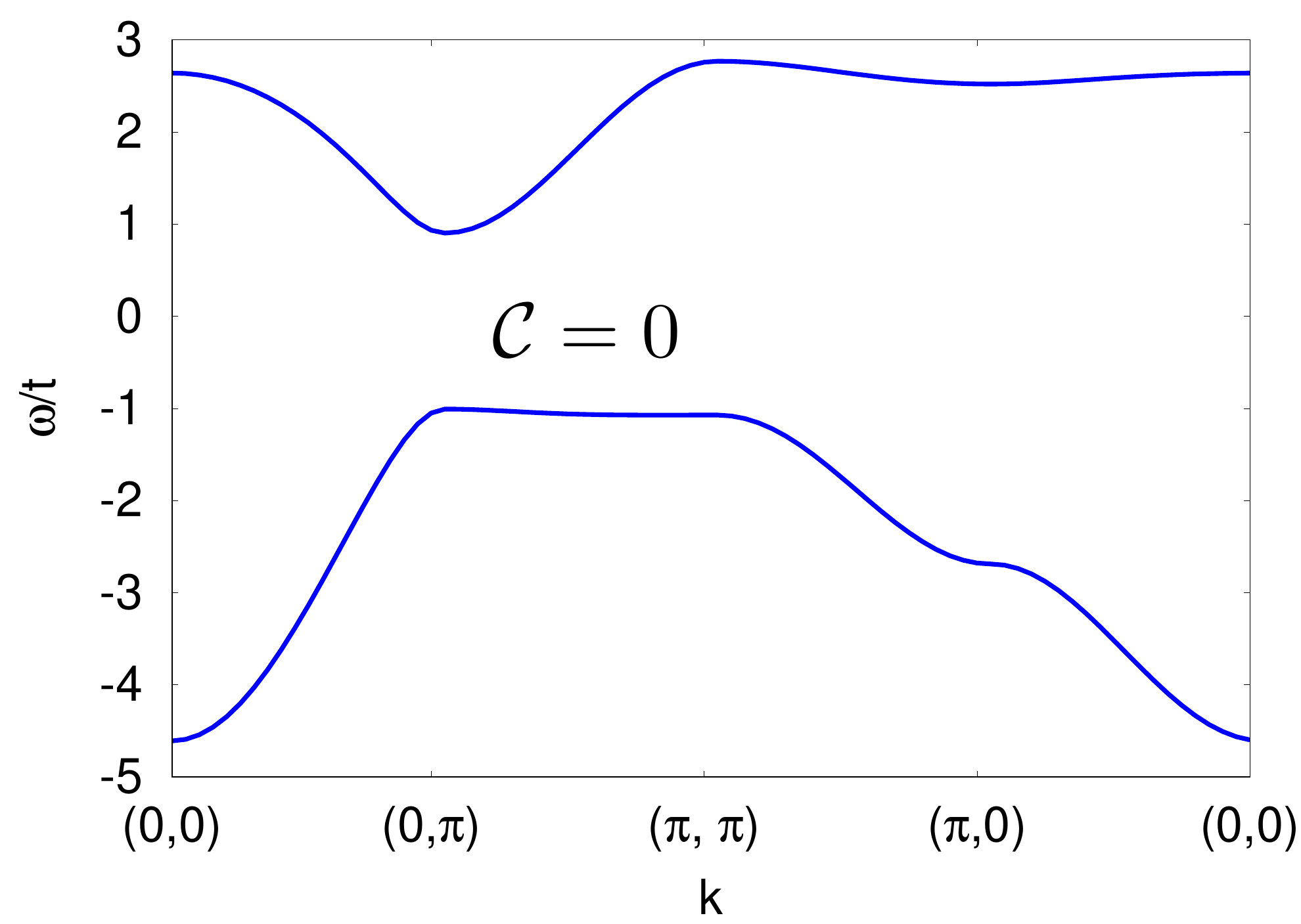}
        \includegraphics[width=0.86\columnwidth]{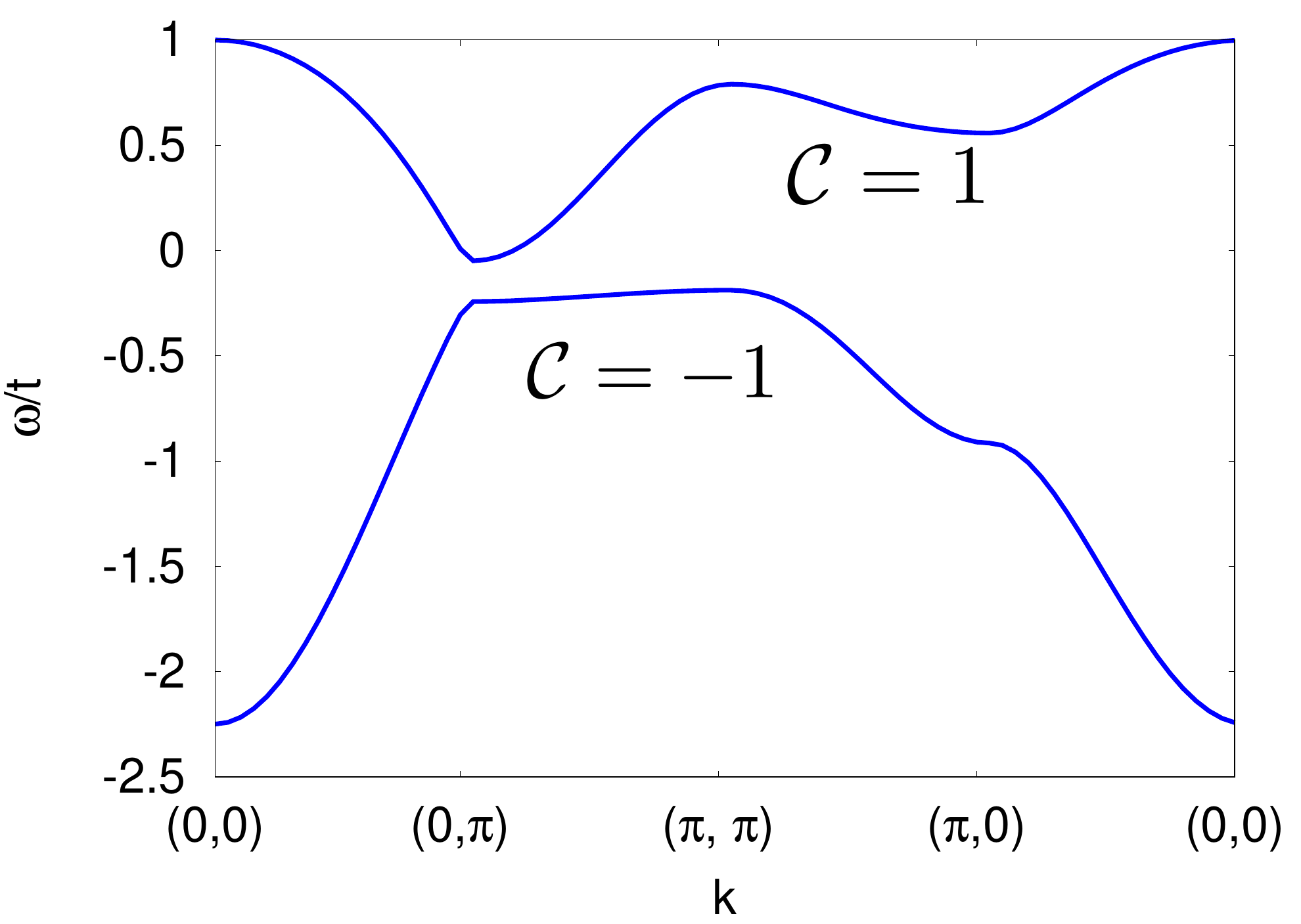}
    \caption{Effective bands calculated with DMFT for $U/t = 8$, $t_c/t = 0.5$, $t^\prime/t = 0.3$, $\delta = 0.9$ at zero temperature (top panel) and at $\beta t = 3.425$ (lower panel).}
    \label{fig:topobands}
\end{figure} 

 Since, the Chern number is inherently related to the presence of Dirac points in the unperturbed spectrum, the absence of those points will yield trivial topological states. Hence, we expect to observe a topological transition by varying the physical parameters such as doping, temperature and interaction strength. For example, in Figure \ref{fig:topobands}  we show the effective bands extracted from our DMFT numerical results for two values of temperature at intermediate coupling: at zero temperature the system is gapped and the bands are topologically trivial, while for a high enough value of temperature the bands carry a non-zero Chern number. 
 
\section{Conclusions}
 We have theoretically and numerically investigated  altermagnetic states arising from a suitably modified Hubbard model.
HF predicts a transition from an altermagnetic metallic state hosting Dirac cones to an insulating gapped state by increasing the Hartree term which is controlled by the order parameter.  Hence, we demonstrated that a transition from  a non-Dirac gapped state to a Dirac magnetic metal can be obtained by doping the system and/or decreasing the interaction strength and/or increasing temperature. 
DMFT confirms the HF predictions at zero temperature for all values of the interaction strength, and at finite temperature for moderate values of $U$.

At finite temperature and at strong coupling, DMFT predicts the emergence of an insulating magnetic state displaying Dirac cones at high-energy: these correlated insulating solutions are characterised by a strong frequency dependence of the self-energy and they cannot be described by a static-mean field ansatz. 
Since the Dirac points tend to be activated by temperature for a wide range of interaction values,  we expect that these states could be stabilised in cold-atom systems where thermal fluctuations play a central role.

In the insulating regime, we analyzed the spin-resolved optical conductivity under finite temperature and strong electronic correlations, focusing on altermagnetic states with varying polarization. Using the Kubo formalism combined with Peierls substitution, we numerically evaluated the conductivity tensor in the strong-coupling limit. Our findings highlight distinct dynamical features: a spin-dependent single-peak structure in the fully polarized altermagnet, and a characteristic double-peak structure with a pronounced minimum in the intermediate polarization regime. This complex behavior appears linked to the emergence of additional high-energy bands,  associated with Dirac cone-like features in the spectral function, illustrating the subtle interplay between electronic correlations and altermagnetic order.
Moreover, we observe that the optical conductivity for spin-up and spin-down electrons is asymmetrically distributed: the spin-up response is shifted toward lower energies, while the spin-down response appears at higher energies relative to the interaction scale $U$. This implies that the two spin channels are activated at different photon energies, suggesting a form of spin-selective optical excitation. Such a mechanism could be exploited for spin-dependent optical filtering or detection, potentially opening avenues for technological applications in spintronic or optoelectronic devices based on correlated materials.

Finally we showed how the Dirac points can be gapped by suitable perturbations to the original Hamiltonian, inducing topological phase transitions as a function of the physical parameters. Given the bulk-edge correspondence \cite{Graf2013}, a topological transition in the bulk will be accompanied by the emergence of edge states on the surface, which can be detected in a cold atomic experiment \cite{Mancini2015,Stuhl2015,Chalopin2020,Braun2024}.

Extension of DMFT such as diagrammatic \cite{rohringer2018} and cluster theories \cite{Maier2005} could be used to   investigate the topological properties of the Green's function zeros when these acquire a non-trivial dependence on the crystalline momentum \cite{blason2023,wagner2023mott,PhysRevB.110.L161106}.

Our results demonstrate that the physics emerging from this model is rich, lying at the intersection of quantum magnetism, strongly correlated systems, Dirac materials \cite{Wehling2014,Vafek2014} and topology. Its realisation represents an ideal target for quantum simulators, particularly in the context of cold atoms trapped in optical lattices. 

 \section{Acknowledgment}
 I thank Andreas Schnyder, Kirill Parshukov, Laura Classen, Giorgio Sangiovanni and Alessandro Toschi for valuable discussions.
\appendix 
\section{Topological invariants}\label{sec:appendix}
Let us consider the following 2$\times$2 Hamiltonian:
\begin{align}
    \mathcal{H}_\sigma(k) &= \alpha_k\mathbb{I}_{2\times 2}+\beta_{k\sigma} \sigma^{(z)} + \gamma_k \sigma^{(x)} +\lambda_{k\sigma} \sigma^{(y)}.
\end{align}
The eigenvalues of this Hamiltonian are given by $\alpha_k\pm \rho_{k\sigma}$, where $\rho_{k\sigma} =\sqrt{\beta_{k\sigma}^{2} + \gamma_{k}^{2}  + \lambda_{k\sigma}^{2} }$ and eigenvectors:
\begin{align}\label{eq:ev}
    \ket{\pm,k,\sigma}&=
    \frac{1}{\sqrt{\rho_{k\sigma}( \rho_{k\sigma} \pm \beta_{k\sigma}) }}\left(
    \begin{array}{c}
         \beta_{k\sigma}\pm \rho_{k\sigma} \\ \gamma_{k} + i\lambda_{k\sigma}
    \end{array}
    \right).
\end{align}
\subsection{Numerical evaluation of the Chern number}
The Chern number for the n-th band can then be calculated as:
\begin{align}
    \mathcal{C}^{(n)}_\sigma &= \frac{1}{2\pi}\sum_k F^{(n)}_{k\sigma},
\end{align}
where:
\begin{align}
    F^{(n)}_{k\sigma} = -\text{Im}\,\text{log}\,U^{(n)}_{k\to k_1,\sigma}U^{(n)}_{k_1\to k_2,\sigma} U^{(n)}_{k_2\to k_3,\sigma}U^{(n)}_{k_3\to k,\sigma},
\end{align}
with $k_1 = k + (2\pi/N,0)$, $k_2 = k + (2\pi/N,2\pi/N)$, $k_3 = k + (0,2\pi/N)$, $U^{(n)}_{k\to k^\prime,\sigma} = \left<n,k,\sigma|n,k^\prime,\sigma\right>$ and $n = \pm$   \cite{Zhao2020,Yamamoto2016} \footnote{Let us note that the equation for $F^{(n)}_{k\sigma}$ differs from the one in Ref.\cite{Zhao2020} by a minus sign because we defined the Berry connection as $A_k =\bra{\psi_k}i\nabla\ket{\psi_k}$.}.

In DMFT, we can calculate the Chern number from the renormalised quasi-particle (QP) Hamiltonian \cite{blason2023}:
\begin{align}\label{eq:eff_hamQP_APP}
{\mathcal{H}}_{k\sigma} &= Z^{1/2}\cdot(\mathcal{H}_{0\sigma}(k) + \Sigma_0) \cdot Z^{1/2}, 
\end{align}
where $\Sigma_0 = \text{diag}(\Sigma_{0\uparrow},\Sigma_{0\downarrow})$ represents the Hartree term, 
$Z = \text{diag}(Z_\uparrow,Z_\downarrow)$ with $Z_{\sigma} = (1-\alpha_\sigma)^{-1}$ being the spin/sub-lattice resolved QP renormalisation factor. Let us now fix the non-interacting Hamiltonian to the perturbed one introduced in the last section in the main text, i.e.

\begin{align}
    \mathcal{H}_{0\sigma}(k) &= \left(
    \begin{array}{cc}
         f_A(k)& \epsilon_k -i\,g_{k\sigma} \\
       \epsilon_k +i\,g_{k\sigma}  & f_B(k)
    \end{array}
    \right).
\end{align}

As mentioned in the last section of the main text, we consider two possible cases: (i) $g_{k\sigma} = -2t_c\cos(\frac{k_x-k_y}{2})$ where the perturbation is spin independent, and (ii) $g_{k\sigma} = -2\sigma t_c\cos(\frac{k_x-k_y}{2})$ , where the sign of the perturbation depends on the spin index.

We observe that even if in single-site DMFT the self-consistence condition is the same in these two cases, the topology of the band structure is sensitive to the spin dependent sign change.
\subsection{Low-energy Hamiltonian}
We can expand the perturbed Hamiltonian for the two different spin species in the proximity of the Dirac points:

\begin{align}
    \mathcal{H}^\pm_{\uparrow} &= v_xq_x \sigma^{(x)}\pm v_zq_y\sigma^{(z)} \pm m_\uparrow\,\sigma^{(y)}, \nonumber \\
    \mathcal{H}^\pm_{\downarrow} &= v_xq_y \sigma^{(x)}\mp v_zq_x\sigma^{(z)} \pm m_\downarrow\,\sigma^{(y)},
\end{align}
where $v_x = 2t \cos(\bar{k}/2)$, $v_y =  -2t^\prime\delta\sin(\bar{k})$,with $\bar{k} = \arccos(\frac{\Delta}{2t^\prime\delta} -1)$, $\mathbf{q} =(k_x-\pi,k_y\mp\bar{k})$ for $\sigma =\uparrow$,  $\mathbf{q} =(k_y\mp\bar{k},k_x-\pi)$  for $\sigma =\downarrow$, and where we omitted the part proportional to the identity matrix. The massive term proportional to $\sigma^{(y)}$ can be spin-dependent or spin independent according to the kind of perturbation added. In case (i) $m_\sigma = m = -2t_c\sin(\bar{k}/2)$, while for case (ii) $m_\sigma = \sigma\, m$.
The  Berry connection for the conduction band is defined as $\mathbf{A}^s(\mathbf{q}) = \bra{+,\mathbf{q},\sigma,s}i\,\nabla\ket{+,\mathbf{q},\sigma,s}$, where we added the index $s = \pm$ that specifies a particular Dirac point.
Hence, the Chern number for the conduction band is given by:
\begin{align}
    \mathcal{C}^+_\sigma &= \sum_s \frac{1}{2\pi }\int d\mathbf{q}\, \partial_x\, A^s_{y}(\mathbf{q})-\partial_y\, A^s_{x}(\mathbf{q}) \nonumber \\ &= \text{sign}(m_\sigma\,v_x\,v_y).
\end{align}

\bibliography{ref_biblio.bib}
\end{document}